\documentclass[twocolumn]{aastex62}
\usepackage{graphicx}
\usepackage{hyperref}
\usepackage{epstopdf}
\usepackage{multirow}
\graphicspath{{./}}

\shorttitle{Broadband variability study of MAXI J1631-479}
\shortauthors{Bu et al.}

\begin{document}

\title{Broadband variability study of MAXI J1631-479 in its Hard-Intermediate State observed with \textit{Insight}-HXMT}

\correspondingauthor{Q.~C. Bu}

\author[0000-0001-5238-3988]{Q.~C. Bu}
\email{buqc@ihep.ac.cn; bu@astro.uni-tuebingen.de}
\affiliation{Key Laboratory of Particle Astrophysics, Institute of High Energy Physics, Chinese Academy of Sciences, 19B Yuquan Road, Beijing 100049, China}
\affiliation{Institut f\"ur Astronomie und Astrophysik, Kepler Center for Astro and Particle Physics, Eberhard Karls Universit\"at, Sand 1, 72076 T\"ubingen, Germany}

\author[0000-0001-5586-1017]{S.~N. Zhang}
\affiliation{Key Laboratory of Particle Astrophysics, Institute of High Energy Physics, Chinese Academy of Sciences, 19B Yuquan Road, Beijing 100049, China}
\affiliation{University of Chinese Academy of Sciences, Chinese Academy of Sciences, Beijing 100049, China}

\author{A. Santangelo}
\affiliation{Institut f\"ur Astronomie und Astrophysik, Kepler Center for Astro and Particle Physics, Eberhard Karls Universit\"at, Sand 1, 72076 T\"ubingen, Germany}

\author{T.~M. Belloni}
\affiliation{INAF-Osservatorio Astronomico di Brera, Via E. Bianchi 46, I-23807 Merate (LC), Italy}

\author[0000-0003-4498-9925]{L. Zhang}
\affiliation{Department of Physics and Astronomy, University of Southampton, Southampton, SO17 1BJ, UK}

\author{J.~L. Qu}
\affiliation{Key Laboratory of Particle Astrophysics, Institute of High Energy Physics, Chinese Academy of Sciences, 19B Yuquan Road, Beijing 100049, China}
\affiliation{University of Chinese Academy of Sciences, Chinese Academy of Sciences, Beijing 100049, China}

\author[0000-0002-2705-4338]{L. Tao}
\affiliation{Key Laboratory of Particle Astrophysics, Institute of High Energy Physics, Chinese Academy of Sciences, 19B Yuquan Road, Beijing 100049, China}

\author{Y. Huang}
\affiliation{Key Laboratory of Particle Astrophysics, Institute of High Energy Physics, Chinese Academy of Sciences, 19B Yuquan Road, Beijing 100049, China}
\affiliation{University of Chinese Academy of Sciences, Chinese Academy of Sciences, Beijing 100049, China}

\author{X. Ma}
\affiliation{Key Laboratory of Particle Astrophysics, Institute of High Energy Physics, Chinese Academy of Sciences, 19B Yuquan Road, Beijing 100049, China}

\author{Z.~S. Li}
\affiliation{Department of Physics, Xiangtan University, Xiangtan 411105, Hunan, China}

\author{S. Zhang}
\affiliation{Key Laboratory of Particle Astrophysics, Institute of High Energy Physics, Chinese Academy of Sciences, 19B Yuquan Road, Beijing 100049, China}

\author{L. Chen}
\affiliation{Department of Astronomy, Beijing Normal University, Beijing 100075, China}

\author{and the Insight-HMXT collaboration: C. Cai}
\affiliation{Key Laboratory of Particle Astrophysics, Institute of High Energy Physics, Chinese Academy of Sciences, 19B Yuquan Road, Beijing 100049, China}
\affiliation{University of Chinese Academy of Sciences, Chinese Academy of Sciences, Beijing 100049, China}

\author{X.~L. Cao}
\affiliation{Key Laboratory of Particle Astrophysics, Institute of High Energy Physics, Chinese Academy of Sciences, 19B Yuquan Road, Beijing 100049, China}

\author{Z. Chang}
\affiliation{Key Laboratory of Particle Astrophysics, Institute of High Energy Physics, Chinese Academy of Sciences, 19B Yuquan Road, Beijing 100049, China}

\author{T.~X. Chen}
\affiliation{Key Laboratory of Particle Astrophysics, Institute of High Energy Physics, Chinese Academy of Sciences, 19B Yuquan Road, Beijing 100049, China}

\author{Y. Chen}
\affiliation{Key Laboratory of Particle Astrophysics, Institute of High Energy Physics, Chinese Academy of Sciences, 19B Yuquan Road, Beijing 100049, China}

\author{Y.~P. Chen}
\affiliation{Key Laboratory of Particle Astrophysics, Institute of High Energy Physics, Chinese Academy of Sciences, 19B Yuquan Road, Beijing 100049, China}

\author{W.~W. Cui}
\affiliation{Key Laboratory of Particle Astrophysics, Institute of High Energy Physics, Chinese Academy of Sciences, 19B Yuquan Road, Beijing 100049, China}

\author{Y.~Y. Du}
\affiliation{Key Laboratory of Particle Astrophysics, Institute of High Energy Physics, Chinese Academy of Sciences, 19B Yuquan Road, Beijing 100049, China}

\author{G.~H. Gao}
\affiliation{Key Laboratory of Particle Astrophysics, Institute of High Energy Physics, Chinese Academy of Sciences, 19B Yuquan Road, Beijing 100049, China}
\affiliation{University of Chinese Academy of Sciences, Chinese Academy of Sciences, Beijing 100049, China}

\author{H. Gao}
\affiliation{Key Laboratory of Particle Astrophysics, Institute of High Energy Physics, Chinese Academy of Sciences, 19B Yuquan Road, Beijing 100049, China}
\affiliation{University of Chinese Academy of Sciences, Chinese Academy of Sciences, Beijing 100049, China}

\author{M.~Y. Ge}
\affiliation{Key Laboratory of Particle Astrophysics, Institute of High Energy Physics, Chinese Academy of Sciences, 19B Yuquan Road, Beijing 100049, China}

\author{Y.~D. Gu}
\affiliation{Key Laboratory of Particle Astrophysics, Institute of High Energy Physics, Chinese Academy of Sciences, 19B Yuquan Road, Beijing 100049, China}

\author{J. Guan}
\affiliation{Key Laboratory of Particle Astrophysics, Institute of High Energy Physics, Chinese Academy of Sciences, 19B Yuquan Road, Beijing 100049, China}

\author{C.~C. Guo}
\affiliation{Key Laboratory of Particle Astrophysics, Institute of High Energy Physics, Chinese Academy of Sciences, 19B Yuquan Road, Beijing 100049, China}
\affiliation{University of Chinese Academy of Sciences, Chinese Academy of Sciences, Beijing 100049, China}

\author{D.~W. Han}
\affiliation{Key Laboratory of Particle Astrophysics, Institute of High Energy Physics, Chinese Academy of Sciences, 19B Yuquan Road, Beijing 100049, China}

\author{J. Huo}
\affiliation{Key Laboratory of Particle Astrophysics, Institute of High Energy Physics, Chinese Academy of Sciences, 19B Yuquan Road, Beijing 100049, China}

\author{S.~M. Jia}
\affiliation{Key Laboratory of Particle Astrophysics, Institute of High Energy Physics, Chinese Academy of Sciences, 19B Yuquan Road, Beijing 100049, China}

\author{W.~C. Jiang}
\affiliation{Key Laboratory of Particle Astrophysics, Institute of High Energy Physics, Chinese Academy of Sciences, 19B Yuquan Road, Beijing 100049, China}

\author{J. Jin}
\affiliation{Key Laboratory of Particle Astrophysics, Institute of High Energy Physics, Chinese Academy of Sciences, 19B Yuquan Road, Beijing 100049, China}

\author{L.~D. Kong}
\affiliation{Key Laboratory of Particle Astrophysics, Institute of High Energy Physics, Chinese Academy of Sciences, 19B Yuquan Road, Beijing 100049, China}
\affiliation{University of Chinese Academy of Sciences, Chinese Academy of Sciences, Beijing 100049, China}

\author{B. Li}
\affiliation{Key Laboratory of Particle Astrophysics, Institute of High Energy Physics, Chinese Academy of Sciences, 19B Yuquan Road, Beijing 100049, China}

\author{C.~K. Li}
\affiliation{Key Laboratory of Particle Astrophysics, Institute of High Energy Physics, Chinese Academy of Sciences, 19B Yuquan Road, Beijing 100049, China}

\author{G. Li}
\affiliation{Key Laboratory of Particle Astrophysics, Institute of High Energy Physics, Chinese Academy of Sciences, 19B Yuquan Road, Beijing 100049, China}

\author{T.~P. Li}
\affiliation{Key Laboratory of Particle Astrophysics, Institute of High Energy Physics, Chinese Academy of Sciences, 19B Yuquan Road, Beijing 100049, China}
\affiliation{University of Chinese Academy of Sciences, Chinese Academy of Sciences, Beijing 100049, China}
\affiliation{Department of Astronomy, Tsinghua University, Beijing 100084, China}

\author{W. Li}
\affiliation{Key Laboratory of Particle Astrophysics, Institute of High Energy Physics, Chinese Academy of Sciences, 19B Yuquan Road, Beijing 100049, China}

\author{X. Li}
\affiliation{Key Laboratory of Particle Astrophysics, Institute of High Energy Physics, Chinese Academy of Sciences, 19B Yuquan Road, Beijing 100049, China}

\author{X.~B. Li}
\affiliation{Key Laboratory of Particle Astrophysics, Institute of High Energy Physics, Chinese Academy of Sciences, 19B Yuquan Road, Beijing 100049, China}

\author{X.~F. Li}
\affiliation{Key Laboratory of Particle Astrophysics, Institute of High Energy Physics, Chinese Academy of Sciences, 19B Yuquan Road, Beijing 100049, China}

\author{Z.~W. Li}
\affiliation{Key Laboratory of Particle Astrophysics, Institute of High Energy Physics, Chinese Academy of Sciences, 19B Yuquan Road, Beijing 100049, China}

\author{X.~H. Liang}
\affiliation{Key Laboratory of Particle Astrophysics, Institute of High Energy Physics, Chinese Academy of Sciences, 19B Yuquan Road, Beijing 100049, China}

\author{J.~Y. Liao}
\affiliation{Key Laboratory of Particle Astrophysics, Institute of High Energy Physics, Chinese Academy of Sciences, 19B Yuquan Road, Beijing 100049, China}

\author{C.~Z. Liu}
\affiliation{Key Laboratory of Particle Astrophysics, Institute of High Energy Physics, Chinese Academy of Sciences, 19B Yuquan Road, Beijing 100049, China}

\author{H.~X. Liu}
\affiliation{Key Laboratory of Particle Astrophysics, Institute of High Energy Physics, Chinese Academy of Sciences, 19B Yuquan Road, Beijing 100049, China}
\affiliation{University of Chinese Academy of Sciences, Chinese Academy of Sciences, Beijing 100049, China}

\author{H.~W. Liu}
\affiliation{Institute of High Energy Physics, Chinese Academy of Sciences, 19B Yuquan Road, Beijing 100049, China}

\author{X.~J. Liu}
\affiliation{Key Laboratory of Particle Astrophysics, Institute of High Energy Physics, Chinese Academy of Sciences, 19B Yuquan Road, Beijing 100049, China}

\author{F.~J. Lu}
\affiliation{Key Laboratory of Particle Astrophysics, Institute of High Energy Physics, Chinese Academy of Sciences, 19B Yuquan Road, Beijing 100049, China}
\affiliation{University of Chinese Academy of Sciences, Chinese Academy of Sciences, Beijing 100049, China}

\author{X.~F. Lu}
\affiliation{Key Laboratory of Particle Astrophysics, Institute of High Energy Physics, Chinese Academy of Sciences, 19B Yuquan Road, Beijing 100049, China}

\author{Q. Luo}
\affiliation{Key Laboratory of Particle Astrophysics, Institute of High Energy Physics, Chinese Academy of Sciences, 19B Yuquan Road, Beijing 100049, China}
\affiliation{University of Chinese Academy of Sciences, Chinese Academy of Sciences, Beijing 100049, China}

\author{T. Luo}
\affiliation{Key Laboratory of Particle Astrophysics, Institute of High Energy Physics, Chinese Academy of Sciences, 19B Yuquan Road, Beijing 100049, China}

\author{R.~C. Ma}
\affiliation{Key Laboratory of Particle Astrophysics, Institute of High Energy Physics, Chinese Academy of Sciences, 19B Yuquan Road, Beijing 100049, China}
\affiliation{University of Chinese Academy of Sciences, Chinese Academy of Sciences, Beijing 100049, China}

\author{B. Meng}
\affiliation{Key Laboratory of Particle Astrophysics, Institute of High Energy Physics, Chinese Academy of Sciences, 19B Yuquan Road, Beijing 100049, China}

\author{Y. Nang}
\affiliation{Key Laboratory of Particle Astrophysics, Institute of High Energy Physics, Chinese Academy of Sciences, 19B Yuquan Road, Beijing 100049, China}
\affiliation{University of Chinese Academy of Sciences, Chinese Academy of Sciences, Beijing 100049, China}

\author{J.~Y. Nie}
\affiliation{Key Laboratory of Particle Astrophysics, Institute of High Energy Physics, Chinese Academy of Sciences, 19B Yuquan Road, Beijing 100049, China}

\author{G. Ou}
\affiliation{Computing Division, Institute of High Energy Physics, Chinese Academy of Sciences, 19B Yuquan Road, Beijing 100049, China}

\author{N. Sai}
\affiliation{Key Laboratory of Particle Astrophysics, Institute of High Energy Physics, Chinese Academy of Sciences, 19B Yuquan Road, Beijing 100049, China}
\affiliation{University of Chinese Academy of Sciences, Chinese Academy of Sciences, Beijing 100049, China}

\author{L.~M. Song}
\affiliation{Key Laboratory of Particle Astrophysics, Institute of High Energy Physics, Chinese Academy of Sciences, 19B Yuquan Road, Beijing 100049, China}
\affiliation{University of Chinese Academy of Sciences, Chinese Academy of Sciences, Beijing 100049, China}

\author{X.~Y. Song}
\affiliation{Key Laboratory of Particle Astrophysics, Institute of High Energy Physics, Chinese Academy of Sciences, 19B Yuquan Road, Beijing 100049, China}

\author{L. Sun}
\affiliation{Key Laboratory of Particle Astrophysics, Institute of High Energy Physics, Chinese Academy of Sciences, 19B Yuquan Road, Beijing 100049, China}

\author{Y. Tan}
\affiliation{Key Laboratory of Particle Astrophysics, Institute of High Energy Physics, Chinese Academy of Sciences, 19B Yuquan Road, Beijing 100049, China}

\author{Y.~L. Tuo}
\affiliation{Key Laboratory of Particle Astrophysics, Institute of High Energy Physics, Chinese Academy of Sciences, 19B Yuquan Road, Beijing 100049, China}
\affiliation{University of Chinese Academy of Sciences, Chinese Academy of Sciences, Beijing 100049, China}

\author{C. Wang}
\affiliation{Key Laboratory of Space Astronomy and Technology, National Astronomical Observatories, Chinese Academy of Sciences, Beijing 100012, China}
\affiliation{University of Chinese Academy of Sciences, Chinese Academy of Sciences, Beijing 100049, China}

\author{L.~J. Wang}
\affiliation{Key Laboratory of Space Astronomy and Technology, National Astronomical Observatories, Chinese Academy of Sciences, Beijing 100012, China}

\author{P.~J. Wang}
\affiliation{Key Laboratory of Space Astronomy and Technology, National Astronomical Observatories, Chinese Academy of Sciences, Beijing 100012, China}
\affiliation{University of Chinese Academy of Sciences, Chinese Academy of Sciences, Beijing 100049, China}

\author{W.~S. Wang}
\affiliation{Computing Division, Institute of High Energy Physics, Chinese Academy of Sciences, 19B Yuquan Road, Beijing 100049, China}

\author{Y.~S. Wang}
\affiliation{Key Laboratory of Particle Astrophysics, Institute of High Energy Physics, Chinese Academy of Sciences, 19B Yuquan Road, Beijing 100049, China}

\author{X.~Y. Wen}
\affiliation{Key Laboratory of Particle Astrophysics, Institute of High Energy Physics, Chinese Academy of Sciences, 19B Yuquan Road, Beijing 100049, China}

\author{B.~Y. Wu}
\affiliation{Key Laboratory of Particle Astrophysics, Institute of High Energy Physics, Chinese Academy of Sciences, 19B Yuquan Road, Beijing 100049, China}
\affiliation{University of Chinese Academy of Sciences, Chinese Academy of Sciences, Beijing 100049, China}

\author{B.~B. Wu}
\affiliation{Key Laboratory of Particle Astrophysics, Institute of High Energy Physics, Chinese Academy of Sciences, 19B Yuquan Road, Beijing 100049, China}

\author{M. Wu}
\affiliation{Key Laboratory of Particle Astrophysics, Institute of High Energy Physics, Chinese Academy of Sciences, 19B Yuquan Road, Beijing 100049, China}

\author{G.~C. Xiao}
\affiliation{Key Laboratory of Particle Astrophysics, Institute of High Energy Physics, Chinese Academy of Sciences, 19B Yuquan Road, Beijing 100049, China}
\affiliation{University of Chinese Academy of Sciences, Chinese Academy of Sciences, Beijing 100049, China}

\author{S. Xiao}
\affiliation{Key Laboratory of Particle Astrophysics, Institute of High Energy Physics, Chinese Academy of Sciences, 19B Yuquan Road, Beijing 100049, China}
\affiliation{University of Chinese Academy of Sciences, Chinese Academy of Sciences, Beijing 100049, China}

\author{S.~L. Xiong}
\affiliation{Key Laboratory of Particle Astrophysics, Institute of High Energy Physics, Chinese Academy of Sciences, 19B Yuquan Road, Beijing 100049, China}

\author{Y.~P. Xu}
\affiliation{Key Laboratory of Particle Astrophysics, Institute of High Energy Physics, Chinese Academy of Sciences, 19B Yuquan Road, Beijing 100049, China}
\affiliation{University of Chinese Academy of Sciences, Chinese Academy of Sciences, Beijing 100049, China}

\author{S. Yang}
\affiliation{Key Laboratory of Particle Astrophysics, Institute of High Energy Physics, Chinese Academy of Sciences, 19B Yuquan Road, Beijing 100049, China}

\author{Y.~J. Yang}
\affiliation{Key Laboratory of Particle Astrophysics, Institute of High Energy Physics, Chinese Academy of Sciences, 19B Yuquan Road, Beijing 100049, China}

\author{Q.~B. Yi}
\affiliation{Key Laboratory of Particle Astrophysics, Institute of High Energy Physics, Chinese Academy of Sciences, 19B Yuquan Road, Beijing 100049, China}
\affiliation{School of Physics and Optoelectronics, Xiangtan University, Xiangtan 411105, Hunan, China}

\author{Q.~Q. Yin}
\affiliation{Key Laboratory of Particle Astrophysics, Institute of High Energy Physics, Chinese Academy of Sciences, 19B Yuquan Road, Beijing 100049, China}

\author{Y. You}
\affiliation{Key Laboratory of Particle Astrophysics, Institute of High Energy Physics, Chinese Academy of Sciences, 19B Yuquan Road, Beijing 100049, China}
\affiliation{University of Chinese Academy of Sciences, Chinese Academy of Sciences, Beijing 100049, China}

\author{F. Zhang}
\affiliation{Institute of High Energy Physics, Chinese Academy of Sciences, 19B Yuquan Road, Beijing 100049, China}

\author{H.~M. Zhang}
\affiliation{Computing Division, Institute of High Energy Physics, Chinese Academy of Sciences, 19B Yuquan Road, Beijing 100049, China}

\author{J. Zhang}
\affiliation{Key Laboratory of Particle Astrophysics, Institute of High Energy Physics, Chinese Academy of Sciences, 19B Yuquan Road, Beijing 100049, China}

\author{P. Zhang}
\affiliation{Key Laboratory of Particle Astrophysics, Institute of High Energy Physics, Chinese Academy of Sciences, 19B Yuquan Road, Beijing 100049, China}

\author{W.~C. Zhang}
\affiliation{Key Laboratory of Particle Astrophysics, Institute of High Energy Physics, Chinese Academy of Sciences, 19B Yuquan Road, Beijing 100049, China}

\author{W. Zhang}
\affiliation{Key Laboratory of Particle Astrophysics, Institute of High Energy Physics, Chinese Academy of Sciences, 19B Yuquan Road, Beijing 100049, China}
\affiliation{University of Chinese Academy of Sciences, Chinese Academy of Sciences, Beijing 100049, China}

\author{Y.~F. Zhang}
\affiliation{Key Laboratory of Particle Astrophysics, Institute of High Energy Physics, Chinese Academy of Sciences, 19B Yuquan Road, Beijing 100049, China}

\author{Y.~H. Zhang}
\affiliation{Key Laboratory of Particle Astrophysics, Institute of High Energy Physics, Chinese Academy of Sciences, 19B Yuquan Road, Beijing 100049, China}
\affiliation{University of Chinese Academy of Sciences, Chinese Academy of Sciences, Beijing 100049, China}

\author{H.~S. Zhao}
\affiliation{Key Laboratory of Particle Astrophysics, Institute of High Energy Physics, Chinese Academy of Sciences, 19B Yuquan Road, Beijing 100049, China}

\author{X.~F. Zhao}
\affiliation{Key Laboratory of Particle Astrophysics, Institute of High Energy Physics, Chinese Academy of Sciences, 19B Yuquan Road, Beijing 100049, China}
\affiliation{University of Chinese Academy of Sciences, Chinese Academy of Sciences, Beijing 100049, China}

\author{S.~J. Zheng}
\affiliation{Key Laboratory of Particle Astrophysics, Institute of High Energy Physics, Chinese Academy of Sciences, 19B Yuquan Road, Beijing 100049, China}

\author{D.~K. Zhou}
\affiliation{Key Laboratory of Particle Astrophysics, Institute of High Energy Physics, Chinese Academy of Sciences, 19B Yuquan Road, Beijing 100049, China}
\affiliation{University of Chinese Academy of Sciences, Chinese Academy of Sciences, Beijing 100049, China}


\begin{abstract}

We report the energy-resolved broadband timing analysis of the black hole X-ray transient MAXI J1631-479 during its 2019 outburst from February 11 to April 9, using data from the \textit{Insight-Hard X-ray Modulation Telescope} (\textit{Insight}-HXMT), which caught the source from its hard intermediate state to the soft state. Thanks to the large effective area of \textit{Insight}-HXMT at high energies, we are able to present the energy dependence of fast variability up to $\sim100$\ keV. Type-C quasi-periodic oscillations (QPOs) with frequency varying between $4.9$\ Hz and $6.5$\ Hz are observed in the 1--100\ keV energy band. While the QPO fractional rms increases with photon energy from 1\ keV to $\sim10$\ keV and remains more or less constant from $\sim10$\ keV to $\sim100$\ keV, the rms of the flat-top noise first increases from 1\ keV to $\sim8$\ keV then drops to less than 0.1\% above $\sim30$\ keV. We suggest that the disappearance of the broadband variability above 30\ keV could be caused by the non-thermal acceleration in the Comptonizing plasma. At the same time, the QPOs could be produced by the precession of either a small-scale jet or a hot inner flow model.

\end{abstract}



\section{Introduction} \label{sec:intro}

Galactic black hole transients (BHTs) are discovered mostly as X-ray transients, most of which are low mass X-ray binaries (LMXBs). A typical outburst of a BHT lasts from months to years, displaying characteristic evolution in its X-ray spectral and timing properties, which are further divided into different spectral states \citep{rem06, bel16}. Low frequency quasi-periodic oscillations (LFQPOs) are commonly observed in BHTs. They are observed as narrow peaks in the Fourier power density spectra (PDS) computed from the fast variable light curves. In BH systems, the LFQPOs are known as type-A, B, and C types \citep{cas05, mot15}, with frequency varying from a few mHz to tens of Hertz. The relation between X-ray variability and spectral state evolution in BHTs has been addressed in plenty of works, hence making fast variability an indicator to trace the spectral states during an outburst \citep{bel11}.

A clear pattern of the X-ray spectral evolution is found in most BHTs, known as the ``q-diagram" \citep{hom01, bel16}. For a typical BHT outburst, the source starts from the low hard state (LHS) and lasts over a wide range of luminosity before it evolves into the hard intermediate state (HIMS). During the LHS, the X-ray spectrum is dominated by a non-thermal emission from an optical/thin corona \citep{don07, gil10, bel16}. The disc component starts becoming important in the X-ray flux during the HIMS. Type-C QPOs characterized by a high amplitude (up to $\sim$ 20\% rms) and simultaneous ``flat-top" noise (FTN) are observed in the LHS and HIMS and their frequencies correlate tightly with the spectral photon index \citep{vig03}. The soft intermediate state (SIMS) is usually recognized by the appearance of type-B QPOs, since the energy spectra from the softest HIMS and the SIMS are indistinguishable (at least below $\sim$ 10\ keV) when their hardness ratios are similar \citep{sti11}. However, for each single source the hardness of the SIMS is always lower than in the HIMS. Type-B QPOs are characterized by a relatively high amplitude (up to $\sim$ 5\% rms) and a weak red noise, thereby making them distinguishable from type-C QPOs \citep{rem02, cas05, ing19}. When the disc component becomes dominant in the energy spectra, the source transitions into the soft state (SS) from the SIMS. Type-A QPOs very rarely appear in the SS and have a very weak rms (a few percent rms). Eventually, the source enters into LHS again after passing through the intermediate states in a reverse order.

MAXI J1631-479 is an X-ray binary transient discovered by the \textit{Monitor of All-sky X-ray Image (MAXI)} \citep{kob18} on 2018 December 21. The spectral properties of MAXI J1631-479 obtained with the \textit{Nuclear Spectroscopic Telescope Array (NuSTAR)} indicate that the source is an accreting black hole discovered in its soft state \citep{miy18}. A soft-to-hard state transition was later observed on 2019 January 23 (MJD 58506) \citep{neg19}, accompanied by clear changes in the X-ray flux and variability properties \citep{eij19}.

Evidence for strong reflection features was detected in the X-ray energy spectra of \textit{NuSTAR} observations in the soft state of the outburst \citep{miy18, xu20}. During the HIMS, PDS with LFQPOs and strong peak noise were observed by the \textit{Neutron Star Interior Composition Explorer, (NICER)} \citep{eij19, rou21}. The detected QPO frequency varies between $\sim4$\ Hz and $\sim10$\ Hz, while the total fractional rms in the 0.1--100\ Hz range remains at $\sim12\%$ (3--12\ keV). The QPO shows hard phase lag of $\sim0.1$\ rad between the $\sim4$--6 and $\sim6$--10\ keV energy bands, which is consistent with the QPO phase lags widely seen in low inclination BHTs system \citep{eij17}. Both the spectral and timing properties strongly suggest MAXI J1631-479 being a black hole X-ray binary.

The \textit{Insight-Hard X-ray Modulation Telescope (Insight}-HXMT) carried out Target of Opportunity (ToO) observations on MAXI J1631-479 following \textit{MAXI}/GSC discovery from 2019 February 11 (MJD 58526), covering two month of the outburst, for a total exposure time of $\sim300$\ ks. There are three main scientific payloads on-board of \textit{Insight}-HXMT: the Low Energy Telescope (LE, 1--15\ keV, 384\ cm$^{2}$, $\sim1$\ ms), the Medium Energy Telescope (ME, 5--30\ keV, 952\ cm$^{2}$, $\sim276$\ $\mu$s) and the High Energy (HE, 20--250\ keV, 5000\ cm$^{2}$, $\sim25$\ $\mu$s) Telescope \citep{zha20, liu20, che20, cao20}. The broadband energy coverage, high time resolution and large effective area of \textit{Insight}-HXMT make it an excellent tool in broadband spectral-timing studies of bright X-ray sources \citep{hua18, che18}.

In this work, we report the broadband timing properties of the \textit{Insight}-HXMT ToO observational campaign of MAXI J1631-479. Thanks to its high statistic in the hard X-ray band ($>30$\ keV), we are able to perform a detailed analysis. In Section 2, we introduce the \textit{Insight}-HXMT observations and our data reduction strategy. In Section 3, we present our timing analysis of these observations. We discuss the implications of our results in Section 4.

\section{Observations and Data Reduction} \label{sec:Obs}

In this work, we make use of the \textit{Insight}-HXMT observations of MAXI J1631-479 from 2019 February 11 (MJD 58526) to 2019 April 9 (MJD 58582), for a total of 300 ks exposure time. The average exposure times of each observation are $\sim1.7$\ ks for LE, $\sim4.5$\ ks for ME and $\sim3.5$\ ks for HE.

We process the original data using the HXMTDAS hpipeline\footnote{http://hxmten.ihep.ac.cn/SoftDoc/501.jhtml}. We screen the data according to the suggested criteria of the good time intervals (GTIs) selection: elevation angle (ELV) larger than $10^{\circ}$; elevation angle for the bright earth larger than $30^{\circ}$; geometric cutoff rigidity (COR) larger than 8\ GeV; offset for the point position smaller than $0.1^{\circ}$; at least 300\ s before and after the South Atlantic Anomaly passage. To avoid the possible contamination from the bright earth and nearby sources, only small field of views (FoVs) are applied.

Light curves are extracted from screened files using the \textsc{helcgen}, \textsc{melcgen} and \textsc{lelcgen} tasks. In Figure \ref{fig:all_lc}, we show the net count rates for each instrument in the right panel. Observations P0214003024 and P0214003029 are excluded because of the short GTIs ($<100$\ s), which leaves us a total of 27 observations.

\subsection{Background Subtraction}

\begin{figure*}
	\includegraphics[scale=.55]{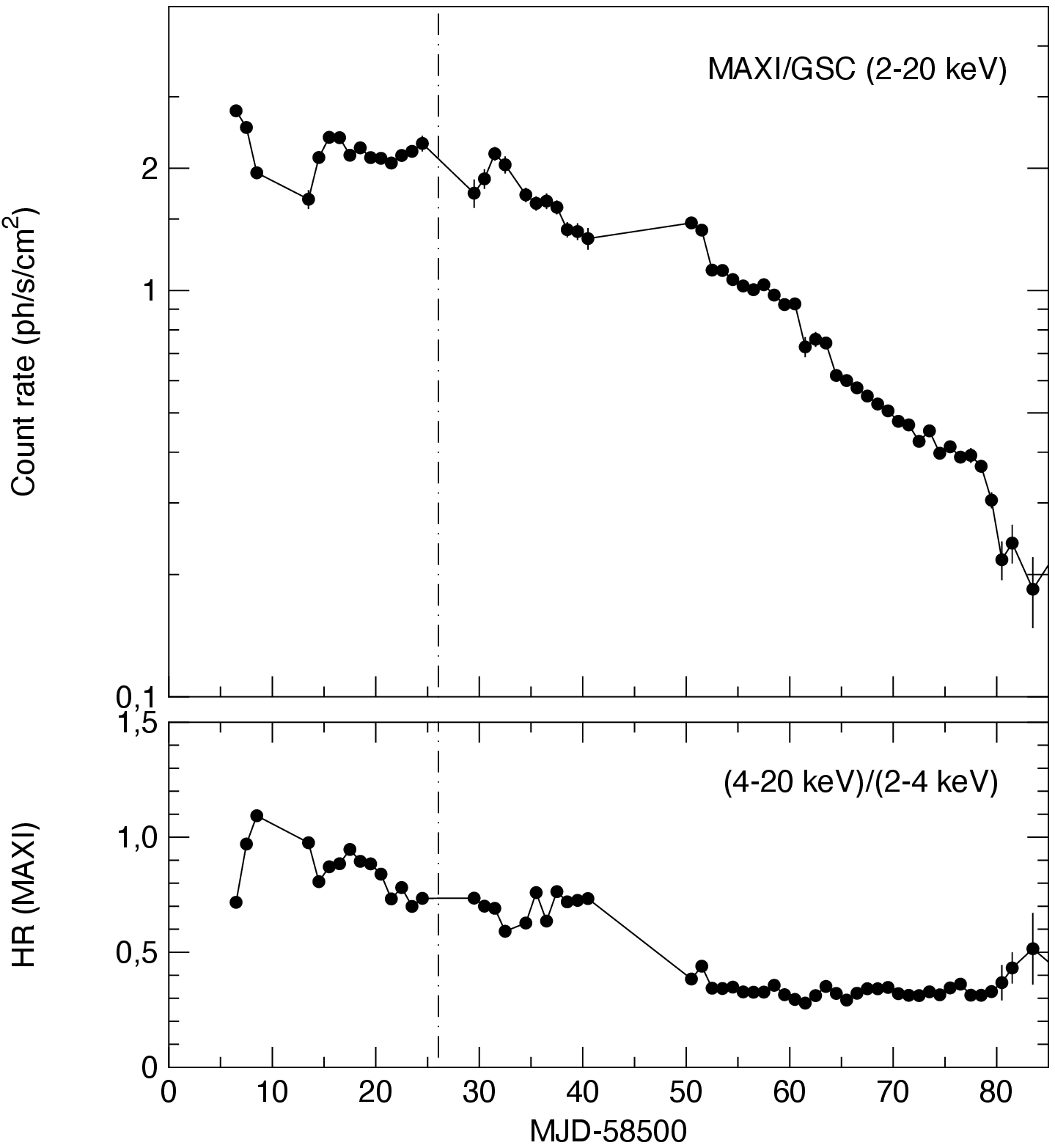}
	\includegraphics[scale=.55]{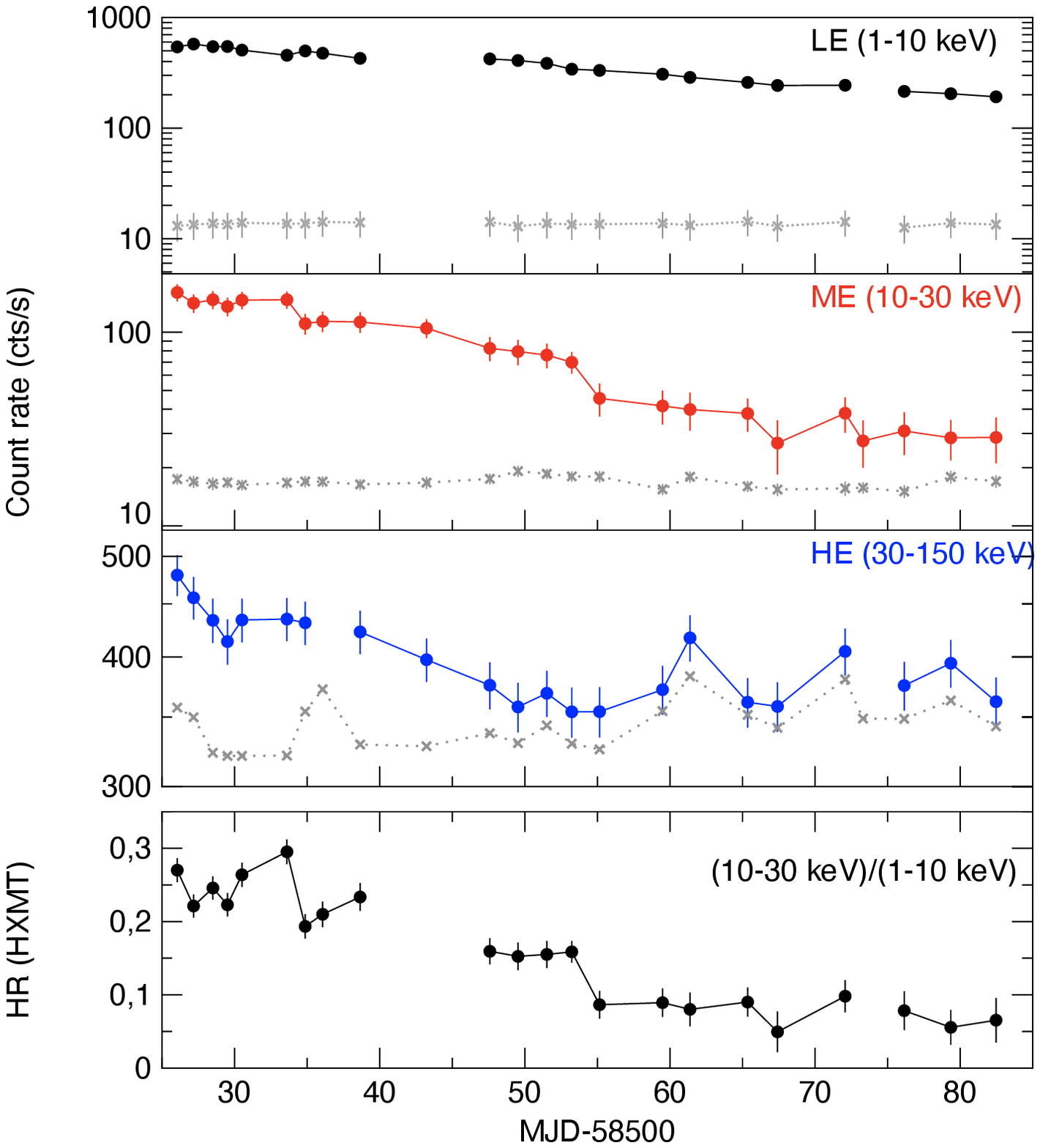}
	\caption{Left panel: (top) long-term \textit{MAXI}/GSC monitoring light curve of the 2018-2019 outburst of MAXI J1631-479. (bottom) Hardness ratio (HR) are estimated as the ratio of count rates in the energy bands of 4--20 and 2--4\ keV of \textit{MAXI}. The data are binned by 24\ h. The dashed line indicates the start time of \textit{Insight}-HXMT observation on the right panel. Right Panel: \textit{Insight}-HXMT light curves and HR of MAXI J1631-479. The gray lines show the background level for the three instruments. HR is estimated as the ratio of net count rates in the energy bands of 10--30 and 1--10\ keV of \textit{Insight}-HXMT. The selected energy bands and instruments are illustrated in the plot. Each point corresponds to one observation. HE data falls below the  detection limit after MJD 58553. \label{fig:all_lc}}
\end{figure*}

The background estimations of HE, ME and LE are performed using the standalone python scripts \textsc{lebkgmap}, \textsc{mebkgmap} and \textsc{hebkgmap} \citep{li20, guo20, liao20a, liao20b}. The averaged background count rate is $\sim10$\ cts/s for LE (1--10\ keV), $\sim35$\ cts/s for ME (10--30\ keV) and $\sim 350$\ cts/s for HE (30--150\ keV). The background level of three telescopes are shown as gray lines in Figure \ref{fig:all_lc}.

It must be noted that there are sources in the field of MAXI J1631-479, as illustrated by the HXMT Bright Source Warning Tool\footnote{http://proposal.ihep.ac.cn/soft/soft2.jspx}, whose contributions can not be modeled by the background software.

Relatively bright contaminating sources included in the small FoVs\footnote{The small FoVs are $1.6^{\circ} \times 6^{\circ}$ for LE, $1^{\circ} \times 4^{\circ}$ for ME and $1.1^{\circ} \times 5.7^{\circ}$ for HE.} of MAXI J1631-479 are 4U 1608-52, GX 339-4, J161741.2-510455 and GX 340+0. 4U 1608-52, GX 339-4 and J161741.2-510455 are transient sources who were in quiescence state during the outburst of MAXI J1631-479 (See the light curves monitored by \textit{MAXI}/GSC, http://maxi.riken.jp/top/slist.html), the background contributions from these sources are thus ignored. GX 340+0 is a persistent Z source and has fast variability, whose X-ray flux stabilizes at $\sim457$\ mCrab in 2--20\ keV (\textit{MAXI}/GSC) and $\sim40$\ mCrab in 15--50\ keV (\textit{Swift}/BAT) during our observation.

\begin{figure*}
\includegraphics[scale=.2]{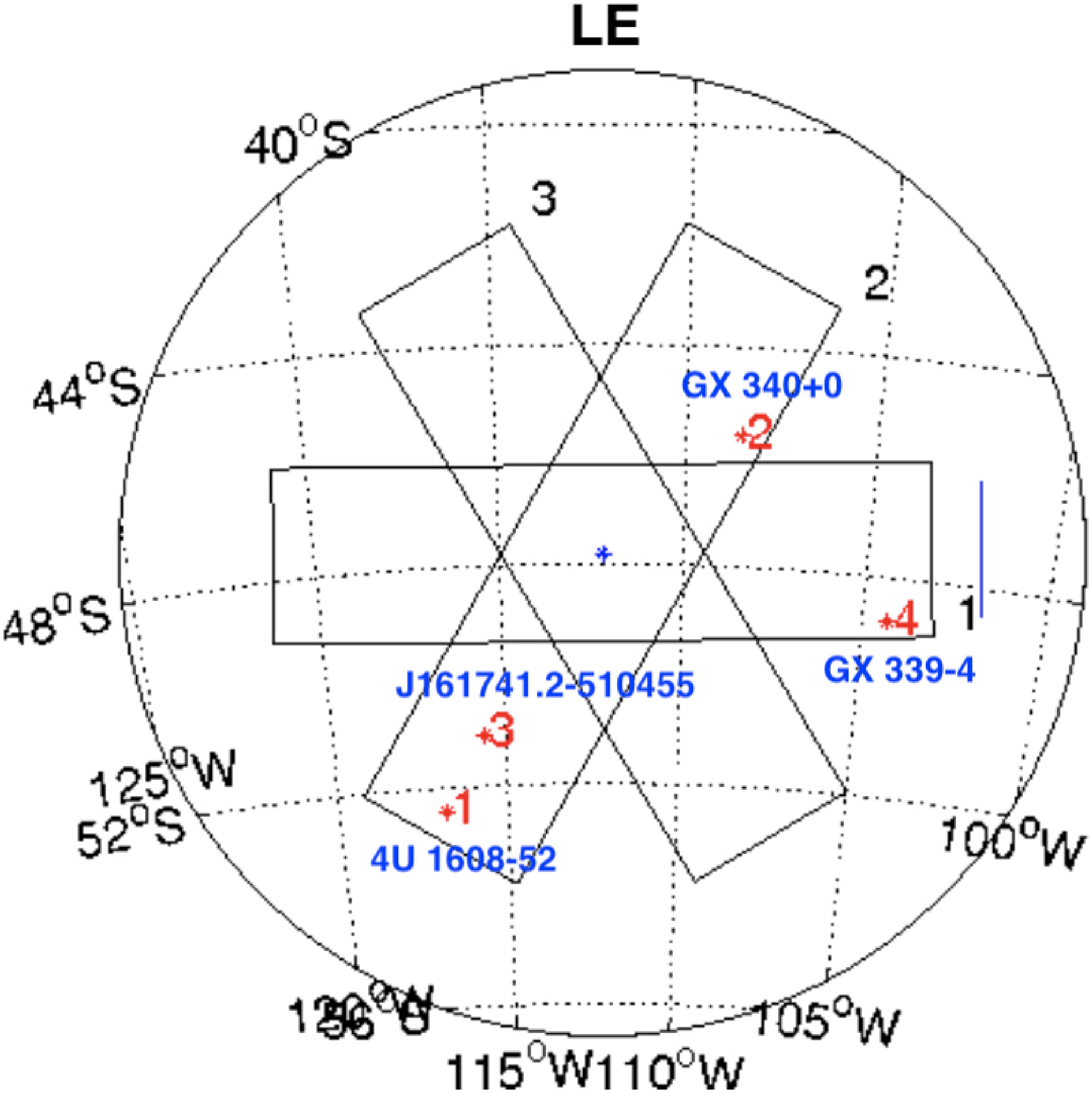}
\includegraphics[scale=.2]{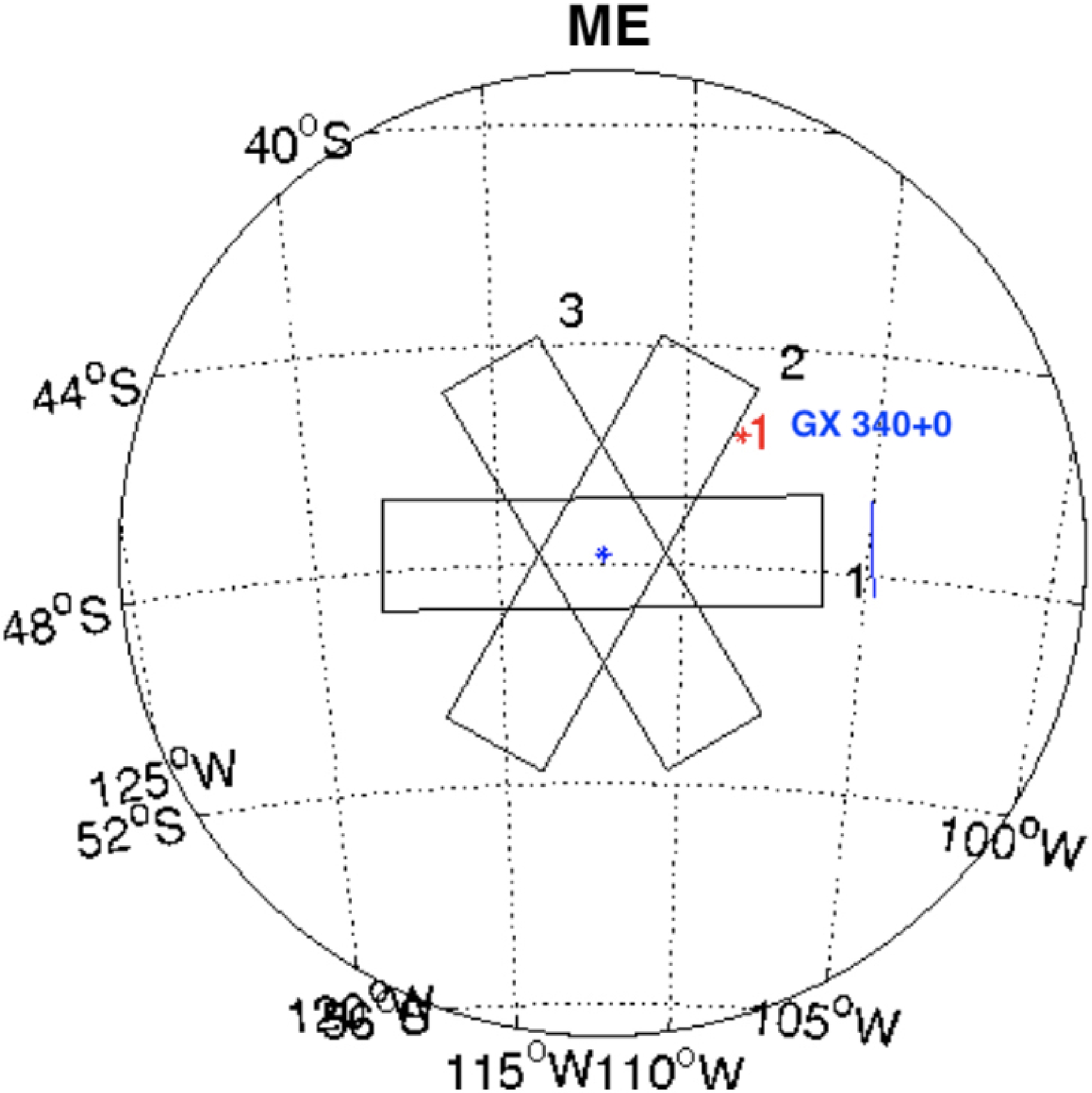}
\includegraphics[scale=.2]{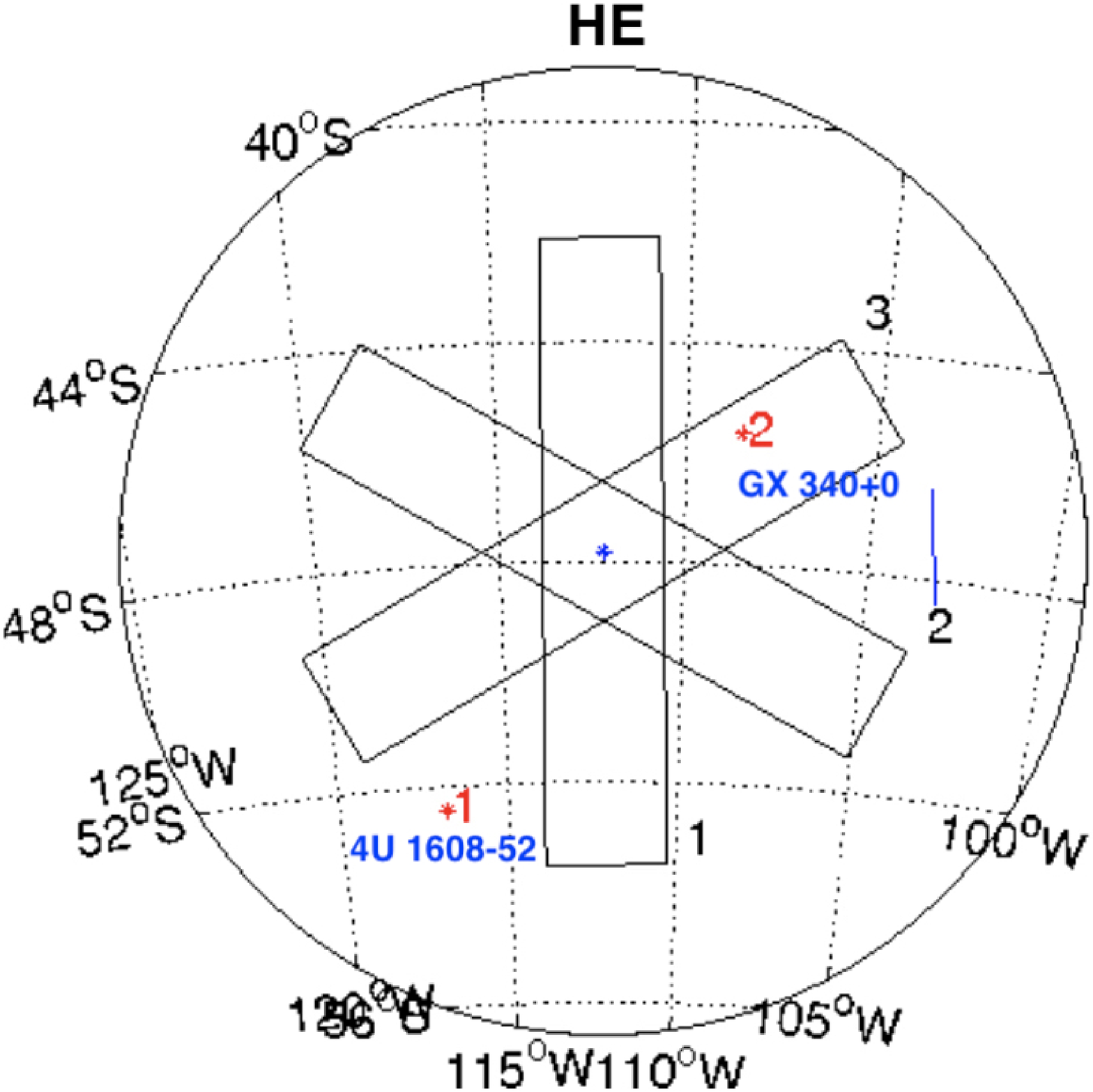}
\caption{The small FoVs of \textit{Insight}-HXMT and the location of contaminating sources compared to our target source during our observation. In order to estimate the contribution of GX 340+0, we create pixel-averaged light curves from each DetBox and compare the flux differences from three telescope. The results are are shown in Figure \ref{fig:pixel}.\label{fig:cal}}
\end{figure*}

\begin{figure}
\includegraphics[width=85mm]{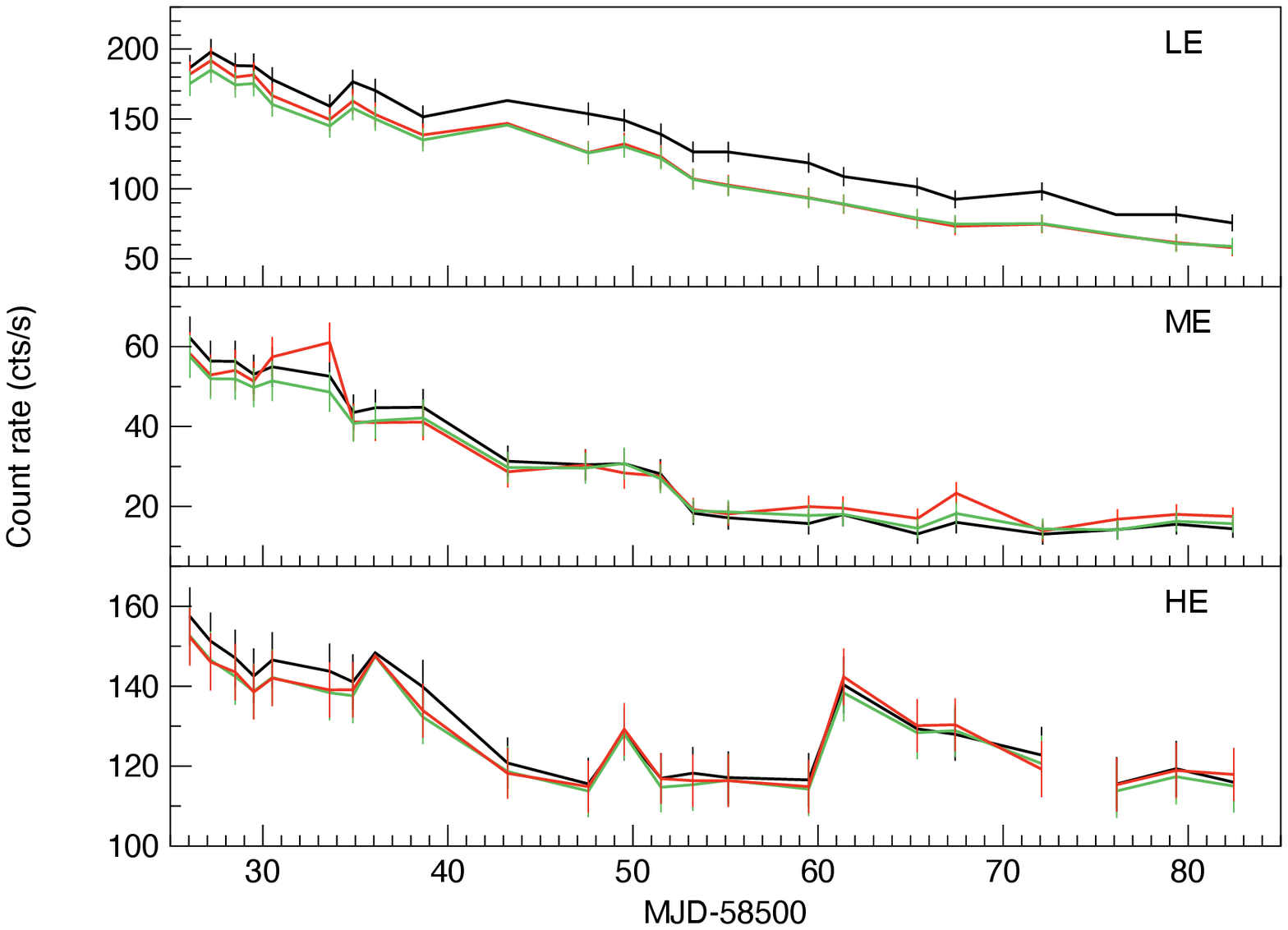}
\caption{The pixel-averaged light curves generated from the three DetBoxes of small FoVs for the three telescopes. The red lines are the ones created from the contaminating DetBox and blue/black lines are from the other two non-contaminated DetBoxes. \label{fig:pixel}}
\end{figure}

The small FoVs of \textit{Insight-HMXT} consists of three DetBoxs (NO. 1, 2, 3) for LE, ME and HE, as shown in Figure \ref{fig:cal}. GX 340+0 constantly appears in the No.2 Detection Box (DetBox) of the LE detector, while occasionally appears in the No.2 DetBox of the ME detector and No.3 DetBox of the HE detector during our observation. In order to estimate the background contribution from GX 340+0, we create pixel-averaged light curves from each DetBox and compare the flux differences between them. The pixel-averaged lightcurve for three telescopes are shown in Figure \ref{fig:pixel}. Our results show that for LE (1--10\ keV), the background contribution from GX 340+0 is less than 1\% and can be ignored; for ME (10--30\ keV), the contribution is $\sim3$\ cts/s and is corrected accordingly for the light curves; for HE (30--150\ keV), the averaged contribution is $\sim1$\ cts/s with a maximum contribution of $\sim5$\%. Considering the systematic error in HE, the HE background contribution from GX 340+0 is negligible.

\section{Analysis and Results}

\subsection{Spectral Evolution} \label{sec:HID}

To address the spectral states evolution during our \textit{Insight}-HXMT observations in a broader context, we plot the long-term \textit{MAXI}/GSC monitoring light curves (Figure \ref{fig:all_lc}, left panel). The \textit{MAXI}/GSC light curve and hardness ratio are produced by the \textit{MAXI}/GSC on-demand web interface\footnote{http://maxi.riken.jp/star\_data/J1631-478/J1631-478.html}. \textit{MAXI}/GSC started the observation of MAXI J1631-479 when the source transitioned to the HIMS on MJD 58506 \citep{neg19, eij19, rou21}. \textit{Insight}-HXMT started the follow-up observation of this source on MJD 58526, 20 days later after the first observation of \textit{MAXI}/GSC while the source was still in the HIMS, which is indicated as the dashed line on the left panel of Figure \ref{fig:all_lc}.

Significant spectral variability is found in the light curves of \textit{Insight}-HXMT), when the count rate in the hard X-ray band ($>10$\ keV) decreases by a factor of $\sim3$ and the soft X-ray band ($<10$\ keV) decreases by a factor of $\sim1.2$. The count rate in the 30--150\ keV becomes undetectable near MJD 58553, suggesting a major reduction from the hard X-ray emission. It is clear that \textit{Insight}-HXMT caught MAXI J1631-479 in the HIMS, when the source shows relatively high variability.

\begin{figure}
\centering
\includegraphics[scale=.45]{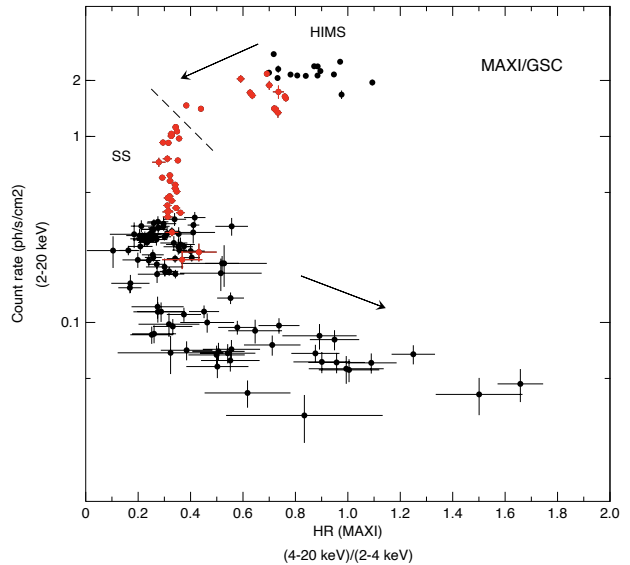}
\caption{The hardness intensity diagram (HID) of MAXI J1631-479 extracted from \textit{MAXI}/GSC. The corresponding applied energy channels are labeled in the plot. The data points are binned by 24\ h. The red points highlight the simultaneous \textit{Insight}-HXMT observations. The dashed line separates the IMS from the SS and the arrows indicate the evolution path of the outburst. \label{fig:maxi_hid}}
\end{figure}

The hardness intensity diagram (HID, also known as ``q-diagram") of Figure \ref{fig:maxi_hid} reports data from \textit{MAXI}/GSC since the source transitioned into the HIMS. In the HID, the hardness ratio is defined between the energy bands 4--20\ keV and 2--4\ keV. The simultaneous \textit{Insight}-HXMT observational period is highlighted in red. Both \textit{MAXI} and \textit{Insight}-HXMT have missed the initial rising phase of the outburst, thus both HIDs are incomplete. In Figure \ref{fig:maxi_hid}, the hardness ratio decreases from $\sim1.1$ to $\sim0.5$ during the first $\sim40$ days, while the source flux (2--20\ keV) varies slightly near $\sim2$\ photons/s/cm$^2$, suggesting that the source is still in the HIMS. Subsequently, the flux drops abruptly from $\sim2.0$ to $\sim0.2$\ photons/s/cm$^2$, while the hardness ratio stays constant at $\sim0.3$, suggesting that the source may have entered the soft state (SS).

\begin{figure}
\includegraphics[scale=.45]{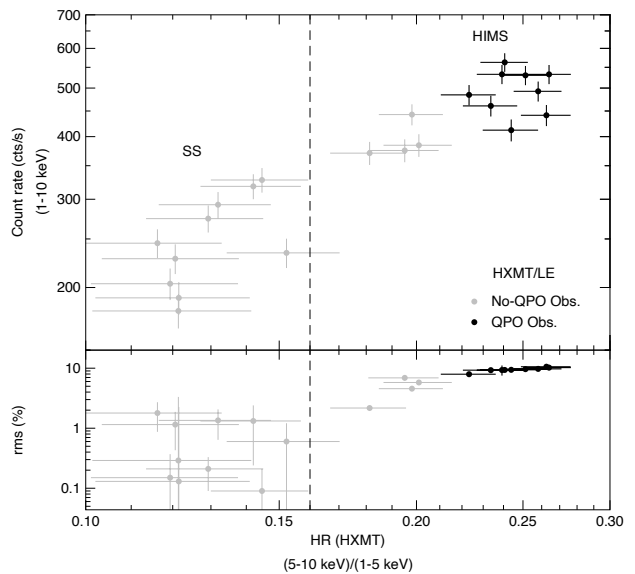}
\caption{Hardness-intensity and hardness-rms diagram produced from the \textit{Insight}-HXMT/LE. The spectral hardness ratio is defined between 5--10\ keV and 1--5\ keV energy bands. The total fractional rms is averaged in frequency range 0.1--32\ Hz for 1--10\ keV. The vertical dashed line separates the HIMS and SS. The QPO observations are highlighted in black. The data are binned by each observation. \label{fig:hid}}
\end{figure}

In order to study the state transition in detail, we calculate the total fractional rms in 1--32\ Hz frequency range for 1--10\ keV using the \textit{Insight}-HXMT/LE observation. In Figure \ref{fig:hid}, we show the HID and hardness-rms diagram (HRD) from the LE. The hardness ratio for LE is defined between the energy bands 5--10\ keV and 1--5\ keV. The QPO observations are highlighted in black. The HID of \textit{HXMT}/LE shows a behavior similar to that of \textit{MAXI}/GSC. In addition, the corresponding total factional rms (1--32\ Hz, 1--10 keV) initially stays at $\sim9.5\%$ for $\sim15$ days and abruptly drops to $4\%$ near MJD 58543, accompanied with a disappearance of QPOs (see section \ref{sec:pds}), which suggests that the source has left the HIMS and entered a softer state. Hereafter, the total rms varies between $4\%$ and $6\%$, while the corresponding hardness ratio remains at $\sim0.2$. During which, the band-limited noise disappears and is replaced by a power-law noise. The total rms drops to less than $1\%$ after MJD 58553 when the emission from $>10$\ keV decreases to less than 10\ cts/s, suggesting that the source may have entered the SS \citep{bel05, hei15}. We therefore give an approximately division of source states based on the HID/HRD properties and mark them as dash lines in the HID/HRD.


\subsection{Power Density Spectra} \label{sec:pds}

For each of the three instruments, we extract an average PDS from each observation by dividing the data into 32\ s intervals and averaging the corresponding PDS. The time resolution is 1\ ms corresponding to a Nyquist frequency of 500\ Hz. The PDS are normalized according to \citet{lea83} and the component due to Poissonian statistics was subtracted according to \citet{zha95}.

The selected energy bands are 1--10\ keV for LE, 10--30\ keV for ME and 30--150\ keV for HE, respectively. The PDS fittings are performed with \textsc{xspec} (12.10.1f), by applying a one-to-one energy-frequency conversion with a unit response. Both broadband noise and QPO peaks in the PDS are fitted with Lorentzian functions \citep{now00, bel02}. We use two Lorentzian peaks for the QPO and its harmonic, and one or two Lorentzian for the broadband noise. Based on the fitting results, we exclude features with a significance\footnote{The significance of QPOs is given as the ratio of the integral of the power of the Lorentzian used to fit the QPO divided by the negative 1$\sigma$ error on the integral of the power.} of less than 3$\sigma$ or a Q factor\footnote{Q = $\nu$/FWHM, where $\nu$ is the centroid frequency of the Lorentzian component and FWHM is its full width at half maximum.} of less than 2.

Among 27 observations, LFQPOs are observed in the first ten observations, as indicated with black dots in the HID of Figure \ref{fig:hid}. The fitting results of these QPOs are given in Table \ref{tab:table1}. Among all the observations, QPOs are observed simultaneously in eight observations by all three telescopes, with their centroid frequencies varying between $\sim4.9$\ Hz and $\sim6.5$\ Hz. While the QPO frequency remains more or less constant with energy, the QPO fractional rms first increases from the LE (1--10\ keV) to ME (10--30\ keV) then slightly decreases from the ME (10--30\ keV) to HE (30--150\ keV). Meanwhile, the total fractional rms increases monotonically with energy from 1 to 150\ keV.

\begin{figure*}
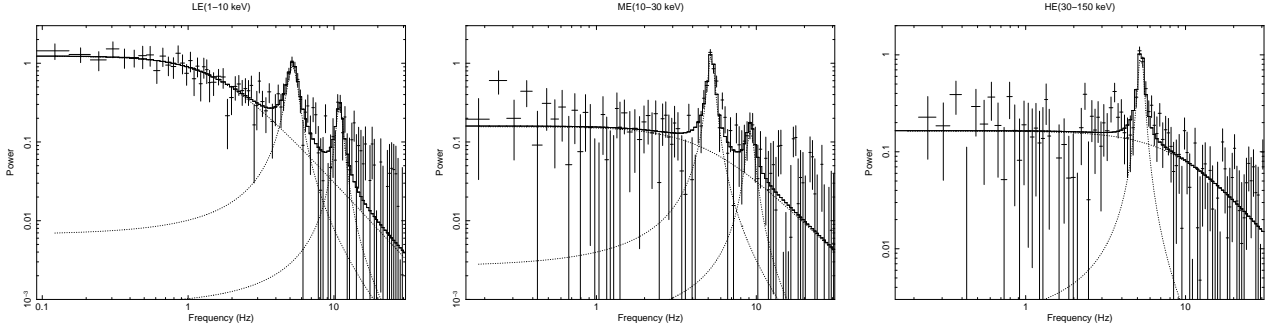

\includegraphics[scale=.23,angle=-90]{006le.eps}
\includegraphics[scale=.23,angle=-90]{006me.eps}
\includegraphics[scale=.23,angle=-90]{006he.eps}
\caption{Power density spectra (PDS) of Observation P0214003006 extracted from 1--10\ keV (left), 10--30\ keV (middle) and 30--150\ keV (right), respectively. A multi-Lorenzian function is applied to fit the spectra. The best-fitting lines are shown in the plots. A QPO signal at $\sim5.2$\ Hz is observed for all three telescopes. The error bars show the 1$\sigma$ level uncertainties. \label{fig:PDS}}
\end{figure*}

All QPOs are observed with a FTN, featuring the characteristics of type-C QPOs, which confirms that these observations correspond to the HIMS. In Figure \ref{fig:PDS}, we show the PDS of ObsID P0214003006\footnote{This observation is selected because of its high count rate and long exposure time.} extracted from the three instruments. The other QPO observations show similar properties, as listed in Table \ref{tab:table1}.

\subsection{Energy-dependent Power Spectra}

\begin{deluxetable}{cc}
\tablecaption{Energy selections for the energy dependence study of MAXI J1631-479 using \textit{Insight}-HXMT. The table shows the selected energy ranges from the three telescopes. \label{tab:table2}}
\tablecolumns{6}
\tablehead{\colhead{Instrument} & \colhead{Energy range (keV)}}
\startdata
LE & 1--5  \\
LE & 5--10 \\
ME & 8--12 \\
ME & 12--20 \\
HE & 28--40 \\
HE & 35--45 \\
HE & 45--95 \\
\enddata
\end{deluxetable}

In order to study energy dependence of the QPO parameters, we divide the 1--100\ keV energy range into seven energy intervals to create the corresponding PDS. Given the good statistic of LE and ME, we divide them into four energy intervals by $\sim$ 5\ keV. Due to the low signal-to-noise ratio of HE, we enlarge the energy bins of HE to get better statistic. The selection of energy ranges are given in Table \ref{tab:table2}.

\begin{figure}
\centering
\includegraphics[width=8cm]{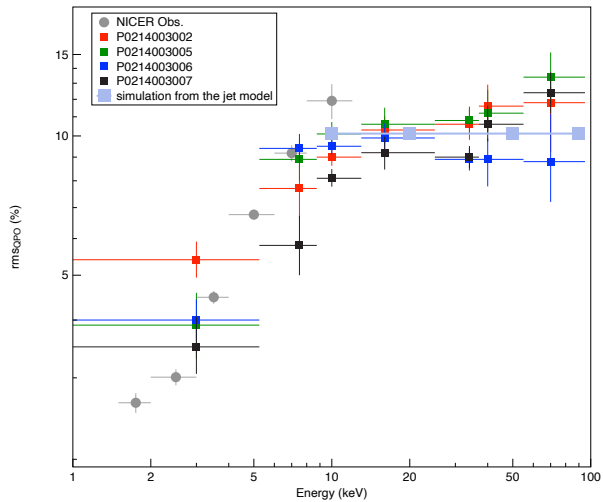}
\caption{The QPO fractional rms as a function of photon energy. The corresponding observation ID are indicated in the plot. The errors are given at 1$\sigma$ level. \label{fig:rms}}
\end{figure}

Among 10 QPO observations, in only four observations (P0214003002, P0214003005, P0214003006 and P0214003007) the QPO is detected in all sub-energy bands with a significance of $>3\sigma$. Figure \ref{fig:rms} shows the QPO fractional rms as a function of energy from 1\ keV to $\sim100$\ keV. It is worth noticing that the last points from the four observations show large error bars, we suggest that the QPO fractional rms remains more or less constant from $\sim10$\ keV to $\sim100$\ keV.

\begin{figure}
\centering
\includegraphics[width=8cm]{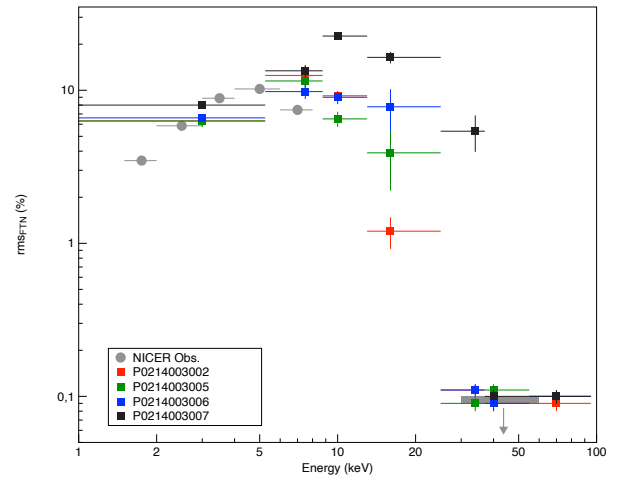}
\caption{The fractional rms of the ``flat-top" noise as a function of photon energy. The corresponding observation IDs are indicated in the plot. The errors are given at 1$\sigma$ level. The grey bar gives an averaged upper limit of the FTN rms. \label{fig:nrms}}
\end{figure}

The fractional rms of the FTN component is plotted in Figure \ref{fig:nrms}. The FTN becomes extremely weak in the high energy band (above 30\ keV). In order to give an upper limit of the FTN rms, we fix the break frequency of the FTN at $\sim$1\ Hz, based on the correlation ($\nu_{\rm{QPO}} =(5.2\pm0.32)*\nu_{\rm{break}}^{(0.81\pm0.03)}$) between LFQPO and break frequency of BHs \citep{bu17, wij99}. The thick grey bar gives an averaged upper limit of the FTN rms given at 1$\sigma$ level. The rms of FTN generally increases with energy from $\sim1$\ keV, reaches its maximum near $\sim8$\ keV and then decreases to $<0.1\%$ above 30\ keV.

In order to compare the rms vs. energy results, we also make use of the data from \textit{NICER} on board the International Space Station (ISS) between 11 Feb and 9 Apr 2019, covering the interval of $Insight$-HXMT observations, for a total of 35 observations. The X-Ray Timing Instrument (XTI) of NICER consists of 56 X-ray optics with silicon detectors and provides high time resolution data in the 0.2--12\ keV energy range \citep{gen12}. We process the original data using the NICERDAS pipeline. For each observation, we extract PDS averaging intervals of 13.1072 seconds over the energy range 0.2--12\ keV. PDS similar to those shown in Figure \ref{fig:PDS} are observed from the first observation until March 3 (MJD 58545). From the next observation, on March 18 (MJD 58560), the PDS are consistent with the SS.

We analyze in more detail the observation with the best statistics, observation 1200500127 on February 11 (MJD 58525). We extract PDS in eight separate energy bands and fit them with a multi-Lorentzian model. Given the high statistics, three harmonically related QPO peaks are seen, in addition to a FTN component and an additional broad-band component at higher frequencies. The QPO main peak is at 4.45\ Hz. No clear signal is detected below 1.5\ keV  despite of the high statistics, and no QPO is detected in the last PDS, 8--12\ keV, due to the lack of sensitivity. The rms vs. energy for the FTN component and main QPO peak are shown as gray points in Figure \ref{fig:nrms} and \ref{fig:rms}. They are compatible with the LE and ME results.

The lag-energy spectra of type-C QPO are also studied with $Insight$-HXMT. Background-subtracted light curves from seven sub-energy bands are generated to calculate the cross-spectra. We extract the complex-valued product of the Fourier transforms used for PDS and produce six cross-spectra using 1--5\ keV as a reference band. The correction for cross-channel talk is included by subtracting the average value of the real part of the cross spectra in 400--500\ Hz frequency range, although this is found to be negligible.

\begin{figure}
\centering
\includegraphics[width=8cm]{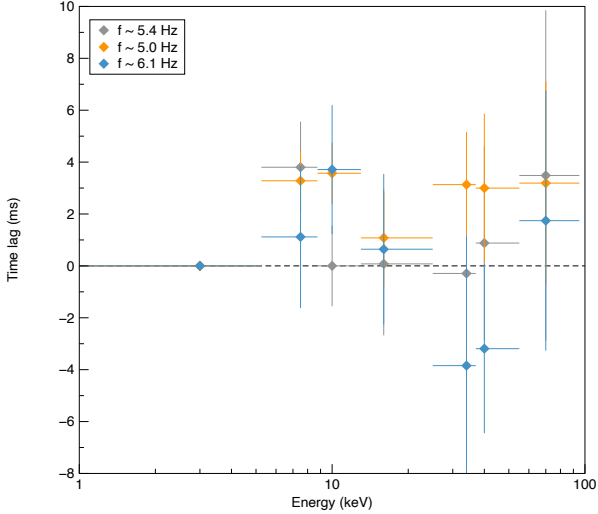}
\caption{The QPO time lag as a function of photon energy. The corresponding QPO centroid frequencies f are indicated in the plot. The error are given at 1$\sigma$ level. \label{fig:lag}}
\end{figure}

The frequency range used for the QPO time lag calculation is the QPO FWHM centered on its centroid frequency. We define hard lag (hard photons lagging soft photons) as positive lag. In order to improve the statistics, we group the QPO observations into three groups according to QPO frequency and average the time lag for each group. The lag-energy spectra for the three groups are shown in Figure \ref{fig:lag}. The QPO time lags generally show hard lags for the whole energy band with a maximum value of $\sim$4\ ms, except for the two points near $\sim$ 40\ keV showing soft lags. However, considering the large errors of the two points, no solid soft lags can be confirmed.

\section{Discussion}

We have reported the energy-resolved timing analysis of the black hole X-ray transient MAXI J1631-479 using data from the \textit{Insight}-HXMT and \textit{NICER}. A state transition from the intermediate state to the soft state is observed near MJD 58553. The source stays for $\sim50$ days in the intermediate state before it enters the soft state, characterized by an abruptly drop in the total variability from $\sim9.5\%$ rms to $\sim1\%$ rms. Type-C QPO with frequency between $4.9$\ Hz and $6.5$\ Hz are observed during the HIMS. While the QPO fractional rms first increases with energy from $\sim1$\ keV to $\sim10$\ keV and remains more or less constant from $\sim10$\ keV to $\sim100$\ keV, the fractional rms of the ``flat-top" noise accompanied with the type-C QPO increases with energy from $\sim1$\ keV to $\sim8$\ keV and then decreases to $<0.1\%$ above 30\ keV. The type-C QPOs generally show hard lags for the whole energy band, with a maximum value of $\sim$4\ ms.

\textit{Insight}-HXMT started the observation of MAXI J1631-479 in its HIMS state where the source showed a high timing variability ($>10\%$ rms) and a presence of type-C QPO. The spectral state evolution is well described by the HID/HRD plot in Figure \ref{fig:maxi_hid} and Figure \ref{fig:hid}, revealing a typical state evolution of a black hole transient from its hard intermediate state to soft state. In proximity of the HIMS/SIMS transition, timing properties constitute the sole way to distinguish between HIMS/SIMS/SS, given the absence of differences in the spectral shape. Generally, the fractional rms decreases during the state transition and is very low in the SS \citep{bel10}, which helps us to distinguish the SS from the intermediate state based on the HRD correlation in Figure \ref{fig:hid}. During the late HIMS, the total fractional rms drops from $>10\%$ to $\sim5\%$ after MJD 58043, accompanied by a replacement of broadband noise to power-law noise and a drop in hardness ratio, which could suggest a transition to the SIMS. However, the lack of type-B QPO during this period makes the identification of SIMS less credible.

The energy spectra of BHTs are generally composed of a soft component associated with an accretion disc and a hard component resulting from Comptonization in a corona. It has been known for years that variability is mainly associated with the hard component rather than the disc. The broadband noise is generally believed to rise from propagation of fluctuations in mass accretion rate produced at different radii \citep{utt05, ing13} and is known to have a soft spectrum in the hard states \citep{gie05}. Similar behavior of broadband noise rms-energy correlation with MAXI J1631-479 (Figure \ref{fig:nrms}) has been found in GX 339-4 when the source was in its late HIMS \citep{bel11}. The broadband noise rms increases during 1--10\ keV and slightly decreases from $\sim$10\ keV to 20\ keV. Comparing to our results from \textit{Insight}-HXMT and \textit{NICER}, we note that the FTN rms already starts to decrease near $\sim8$\ keV and drops to $<0.1\%$ above 30\ keV. These patterns can be explained by considering little variability when disc dominates below $\sim5$\ keV. Above 5\ keV, where the variable power-law dominates, the rms values increase and remain high. When the energy is higher than 8\ keV, the variability of the broadband noise decreases until reaches the extremely low rms values ($< 0.1\%$) in 30--100\ keV band, which can be explained by the effect of non-thermal acceleration in the hot plasma. This effect is dominated by the Compton cooling, the electrons in the hard state are efficiently thermalized even when the power provided to them is entirely in the form of non-thermal acceleration \citep{zdz90, cop99}. In this scenario, a varying seed photon temperature and optical depth of the Comptonizing plasma could lead to the decrease of variability from $\sim10$ to $\sim100$\ keV \citep{gie05}.

Although the physical origin of type-C QPOs is still under debate, many pieces of evidence have suggested a geometric origin \citep{gil10}. Type-C QPOs are particularly prominent in the HIMS of BHTs and are widely believed to be resulted from the Lense-Thirring (LT) precession of a radially extended section of the hot inner flow \citep{ste98, ing09}. This geometric origin of type-C QPOs is further supported by the inclination dependence studies of the QPO characteristics, i.e., the correlations between the QPO rms \citep{sch06, mot15}, the sign of the energy-dependent lags \citep{eij17} and the orbital inclination.

\begin{figure*}
\centering
\includegraphics[width=8cm]{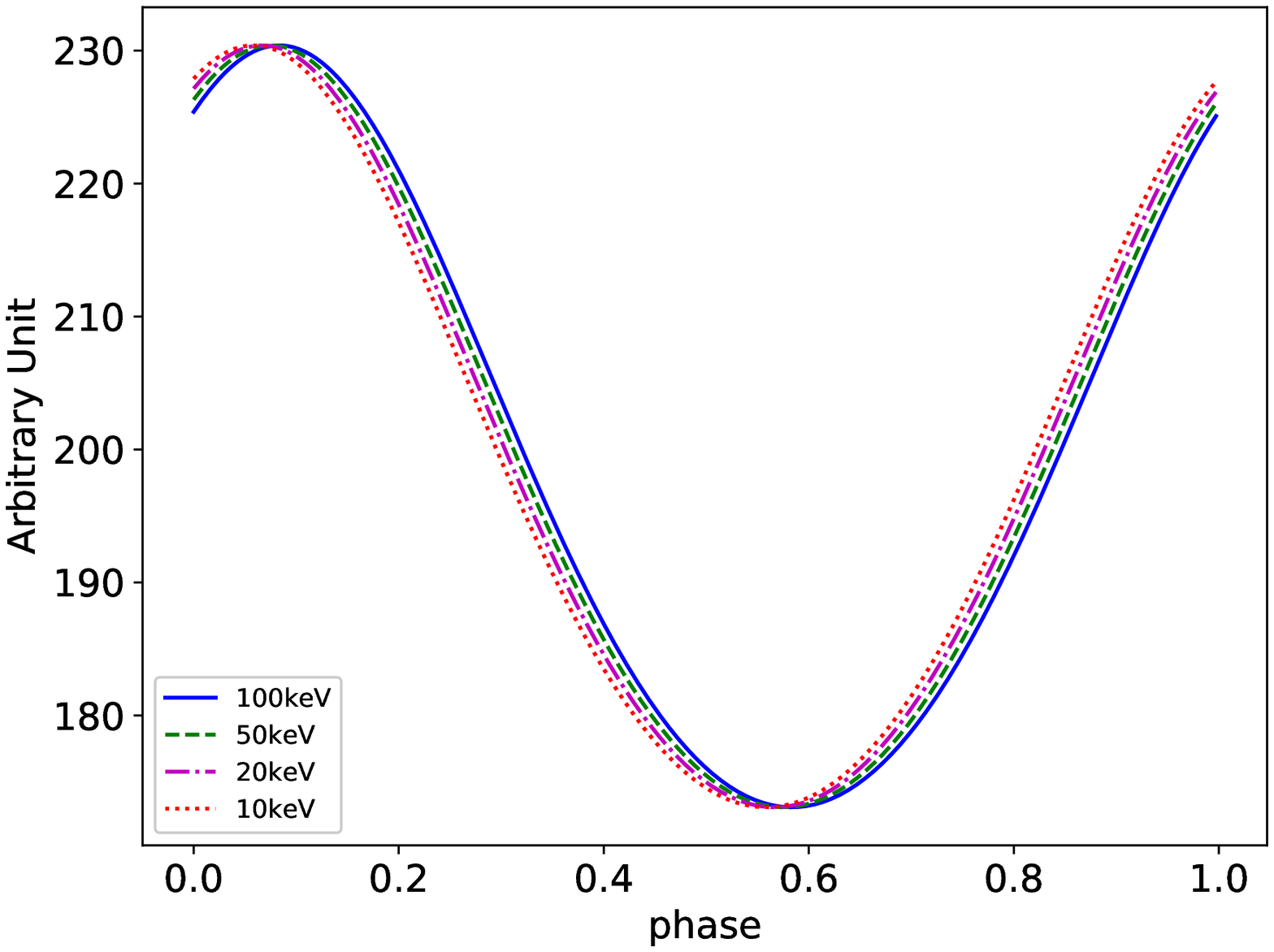}
\includegraphics[width=8.5cm]{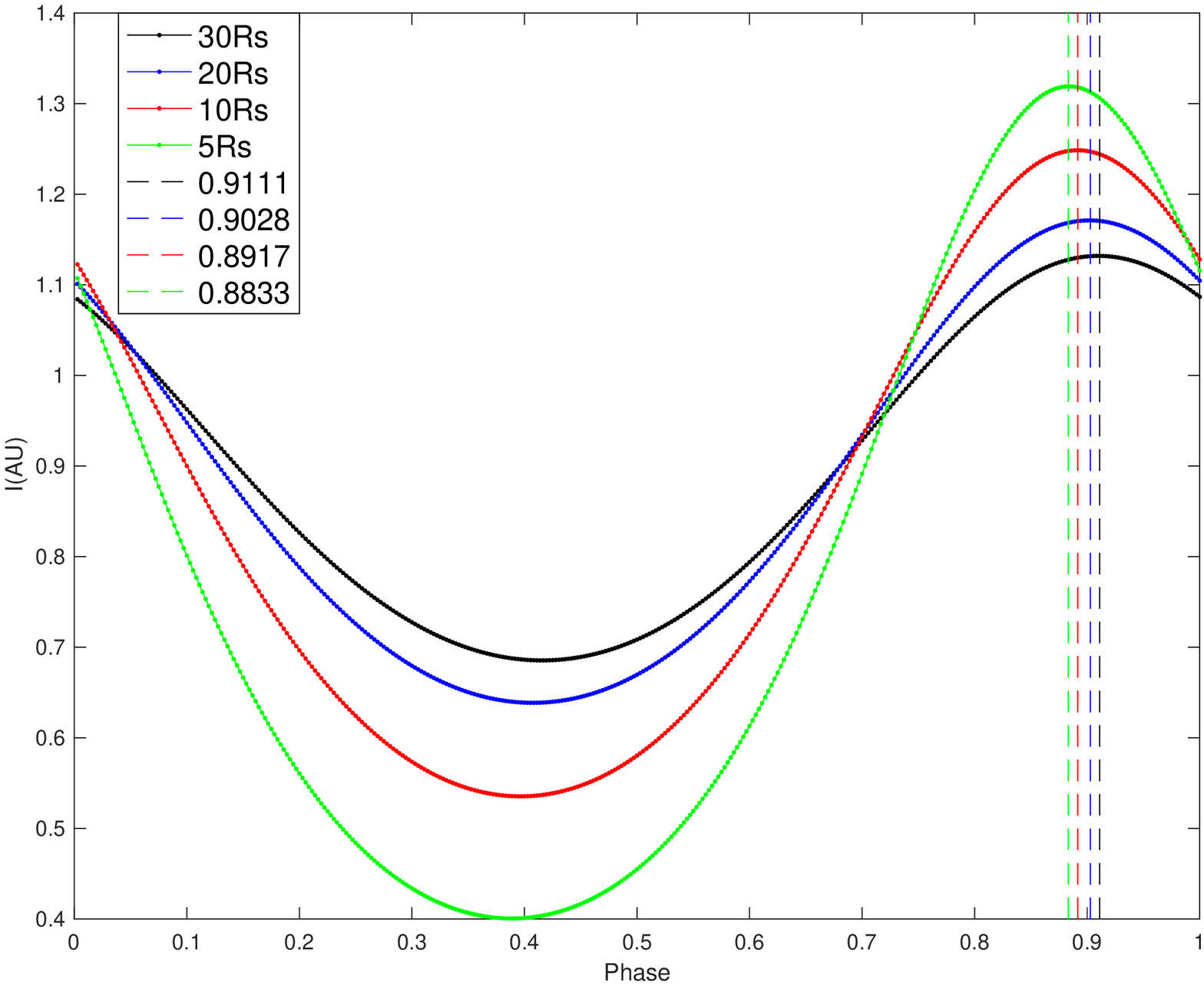}
\caption{Simulated light curve from LT-jet model (left) and LT-hot flow model (right). Assuming an inclination angle of $29\pm1^{\circ}$, a misalignment angle of $10^{\circ}$ for both models. Different colors show simulated flux from different energy bands, while the phase differences among these waveforms give the simulated phase lags. In both models, the observed fluxes are modulated by the Doppler boosting and solid angle effect. (Left) The fluxes are simulated by the observed intensity from different energy bands. (Right) The fluxes are simulated for the hot flow extending from 5 to 30 Rg, which approximately represents the fluxes from 100 to 10\ keV. The vertical lines indicate the peak phase values for the four QPO waveforms.\label{fig:models}}
\end{figure*}

The type-C QPO rms-energy correlation in 2--30\ keV has been extensively studied with Rossi-XTE for the last two decades, while similar energy dependence relations are found in GRS 1915+015 \citep{rod04, yan12, yad16}, H1743-322 \citep{li13a}, XTE J1859+226 \citep{cas04}, and XTE J1550-564 \citep{li13b}. The QPO rms in these sources are found to increase with energy below 10\ keV and become flat above 10\ keV, which are similar to the energy dependence relation we found in MAXI J1631-479. \citet{you18} simulated the fractional rms spectra of the type-C QPOs under the framework of LT precession model and suggested that the flattening above 10\ keV is caused by the high orbital inclination angle. However, both \textit{NuSTAR} spectral analysis \citep{xu20} and \textit{NICER} time lag studies \citep{eij19} suggest that MAXI J1631-479 is likely to be a low inclination system, who should have an increasing QPO rms with energy as predicated by \citet{you18}. Our result shows that the QPO rms increases from 1 to 10\ keV and becomes flat from 10 to 100\ keV, which is inconsistent with the prediction of \citet{you18}.

Thanks to the large effective area of \textit{Insight}-HXMT at high energies, we are able to extend the QPO study up to hundreds of keV, which has rarely been explored by previous satellites. Similar rms energy dependence relation above 30\ keV (see Figure \ref{fig:rms}) has been found in GRS 1915+105 with Rossi-XTE \citep{tom01}, MAXI J1535-571 \citep{hua18} and MAXI J1820+070 \citep{ma20} with \textit{Insight}-HXMT, in which geometric origins of type-C QPO are suggested. The large soft lag ($\sim0.9$\ s) above 200\ keV and energy-related behaviours of the type-C QPOs found in MAXI J1820+070 have posed a great challenge for the existing models \citep{ma20}. \citet{ma20} proposed a model based on the LT precession of a small-scale jet to describe the origin of type-C QPOs, in which the jet twists and rotates around the BH spin axis and the Doppler boosting and solid angle effect modulate the observed flux. In this process, LFQPOs at different energies would be produced from the different parts of the jet with the same frequency. The rms of the LFQPO (i.e., the amplitude of variability in the observed flux) depends on the jet speed (v) and the jet projected angle along the line-of-sight ($\theta$). The phase lag solely depends on the difference in the projected jet angle on the X-Y plane that is perpendicular to the BH spin axis.

We apply the LT precession of the jet model to explain our results (energy dependence of QPO rms and time lag) from MAXI J1631-479. Assuming a orbital inclination angle of $29\pm1^{\circ}$ \citep{xu20}, a misalignment angle of $10^\circ$ between the BH spin and a jet speed of 0.85-0.9c, the QPO fractional rms of $\sim10\%$ above 10\ keV can be reproduced (see blue squares in Figure \ref{fig:rms}). When the energy is below 10\ keV, both the disc and hard components contribute to the variability, thus the QPO rms can not be accurately predicted by the jet model. The time lag $\sim$4\ ms can also be reproduced if considering a small difference in the projected jet angles ($\sim3^\circ$, see left panel in Figure \ref{fig:models}) on the X-Y plane that is perpendicular to the BH spin axis, which suggests that the jet is slightly curved. The jet model can also explain the timing and spectral evolution as the jet size and velocity changes. The QPO frequency increases when the height of jet decreases, meanwhile the spectra soften since more hard photons from the jet are reflected by the accretion disc. When the source evolves to the late HIMS, the QPO rms decreases, which could be explained by the change of the jet speed.

In order to make a comparison, we also apply the LT precession of the truncated disk hot flow model to our results, in which the hot flow precesses as a solid body extending from an inner radius to the truncation radius of the cold outer disk. The observed flux is modulated by the Doppler boosting and solid angle effect to the observer. Assuming an orbital inclination angle of $29\pm1^{\circ}$, an inner radius of 5 Schwarzschild radius (Rs) and a truncation radius of 30\ $Rs$ for the hard state, the QPO fractional rms of $\sim10\%$ above 10\ keV and QPO phase lag of $\sim$0.1--0.2 between 1-5\ keV and 45-95\ keV can be reproduced (see right panel in Figure \ref{fig:models}). Different from the jet model, the phase lag depends on the observer's azimuth and inclination angles.


\acknowledgments
We are grateful for the anonymous referee's helpful comments and suggestions. This research has made use of the data from the \textit{Insight}-HXMT mission, a project funded by China National Space Administration (CNSA) and the Chinese Academy of Sciences (CAS), and data and/or software provided by the High Energy Astrophysics Science Archive Research Center (HEASARC), a service of the Astrophysics Science Division at NASA/GSFC. A/GSFC. This work has also made use of the MAXI data provided by RIKEN, JAXA and the MAXI team. Q.C.B. thanks support from the China Postdoctoral Science Foundation, the National Program on Key Research and Development Project (Grant No. 2016YFA0400801) and the National Natural Science Foundation of China (Grant No. U1838201, U1838202, 11733009, 11673023, U1838111, U1838108 and U1938102). T.M.B. acknowledges financial contribution from the agreement ASI-INAF n.2017-14-H.0. and PRIN-INAF 2019 n.15. L.Z. acknowledges support from the Royal Society Newton Funds. Z.S.L. acknowledges support from the National Natural Science Foundation of China (U1938107). T.L. acknowledges support from the National Natural Science Foundation of China (U1838115).

\begin{deluxetable*}{ccccccc}
	\tablecaption{Best-fitting parameters of the QPO observations of MAXI J1631-479 extracted from \textit{Insight}-HXMT. The table shows the observation starting time\ (MJD), effective exposure time\ (Exp), total fractional rms (rms$_{\rm{total}}$) averaged in the 0.1--32\ Hz frequency band, QPO centroid frequency ($\nu$), QPO full width at half maximum (FWHM) and QPO fractional rms (rms$_{\rm{QPO}}$) for three telescopes. The errors are estimated with 1$\sigma$ level uncertainties. The energy bands selected for study are 1--10\ keV for the LE, 10--30\ keV for the ME and 30--150\ keV for the HE, respectively. The details on the extraction of the power density spectra and on the fitting procedures are described in Section \ref{sec:pds}. \label{tab:table1}}
	\centering
	\tablehead{\colhead{ID} & \colhead{MJD} & \colhead{Exp\ (ks)} & \colhead{rms$_{\rm{total}}$} & \colhead{ $\nu$\ (Hz)} & \colhead{FWHM\ (Hz)} & \colhead{rms$_{\rm{QPO}}$}}
	\startdata
	& & &  LE (1--10\ keV) \\ \hline
	P0214003002 & 58526.0 & 0.9 & 10.2$\pm$0.3 & 5.0$\pm$0.1 & 1.0$\pm$0.2 & 5.6$\pm$0.4 \\
	P0214003003 & 58527.1 & 0.8 & 9.4$\pm$0.3 & 6.1$\pm$0.1 & 1.1$_{-0.3}^{+0.4}$ & 4.7$_{-0.4}^{+0.5}$ \\
	P0214003004 & 58528.4 & 1.4 & 9.6$\pm$0.3 & 5.4$\pm$0.1 & 1.4$\pm$0.3 & 5.3$\pm$0.4 \\
	P0214003005 & 58529.4 & 1.6 & 9.3$_{-0.2}^{+0.4}$ & 6.2$\pm$0.1 & 0.9$\pm$0.3 & 4.1$\pm$0.5 \\
	P0214003006 & 58530.4 & 1.9 & 9.7$\pm$0.2 & 5.2$\pm$0.1 & 0.9$_{-0.1}^{+0.2}$ & 5.1$\pm$0.3 \\
	P0214003007 & 58533.5 & 2.1 & 10.6$\pm$0.2 & 4.9$\pm$0.1 & 0.9$\pm$0.2 & 4.8$\pm$0.3 \\
	P0214003008 & 58534.8 & 2.7 & 8.0$\pm$0.2 & 6.9$\pm$0.1 & 1.8$_{-0.5}^{+0.6}$ & 4.0$\pm$0.4 \\
	P0214003009 & 58536.0 & 1.3 & 9.3$\pm$0.3 & 6.1$\pm$0.1 & 0.5$_{-0.3}^{+0.4}$ & 2.6$\pm$0.6 \\
	P0214003010 & 58538.6 & 2.5 & 9.4$\pm$0.2 & 5.4$\pm$0.1 & 0.5$_{-0.1}^{+0.2}$ & 3.6$\pm$0.4 \\
	P0214003011 & 58541.2 & - & - & - & - & - \\ \hline
	& & & ME (10--30\ keV) \\\hline
	P0214003002 & 58526.0 & 3.1 & 13.2$_{-0.5}^{+0.4}$ & 5.0$\pm$0.1 & 0.9$\pm$0.1 & 10.0$_{-0.4}^{+0.5}$ \\
	P0214003003 & 58527.1 & 2.0 & 14.8$_{-0.7}^{+0.6}$ & 6.1$_{-0.02}^{+0.1}$ & 0.9$\pm$0.2 & 9.6$\pm$0.6 \\
	P0214003004 & 58528.4 & 3.2 & 13.8$\pm$0.6 & 5.5$\pm$0.04 & 0.9$\pm$0.1 & 9.6$\pm$0.5 \\
	P0214003005 & 58529.4 & 3.3 & 13.8$_{-0.6}^{+0.5}$ & 6.2$\pm$0.04 & 0.7$\pm$0.1 & 8.7$\pm$0.5 \\
	P0214003006 & 58530.4 & 3.2 & 13.6$_{-0.6}^{+0.5}$ & 5.2$_{-0.02}^{+0.01}$ & 0.4$\pm$0.1 & 8.4$\pm$0.4 \\
	P0214003007 & 58533.5 & 5.5 & 17.9$_{-0.4}^{+0.3}$ & 4.8$_{-0.04}^{+0.05}$ & 0.6$\pm$0.1 & 7.6$_{-0.5}^{+0.4}$ \\
	P0214003008 & 58534.8 & 8.6 & 13.8$_{-0.5}^{+0.4}$ & 6.6$\pm$0.1 & 2.8$_{-0.8}^{+0.2}$ & 13.0$_{-2.2}^{+1.1}$ \\
	P0214003009 & 58536.0 & 6.6 & 13.5$_{-0.5}^{+0.4}$ & 6.2$\pm$0.04& 0.9$\pm$0.1 & 9.5$_{-0.4}^{+0.5}$ \\
	P0214003010 & 58538.6 & 4.9 & 11.8$_{-0.7}^{+0.6}$ & 5.4$\pm$0.1 & 0.8$\pm$0.1 & 9.1$\pm$0.5 \\
	P0214003011 & 58541.2 & 4.2 & 14.3$_{-0.3}^{+0.6}$ & 6.2$\pm$0.1 & 1.5$_{-0.2}^{+0.3}$ & 10.0$_{-0.3}^{+0.5}$ \\ \hline
	& & & HE (30--150\ keV) \\ \hline
	P0214003002 & 58526.1 & 2.0 & 15.8$\pm$0.6 & 5.3$\pm$0.04 & 0.6$\pm$0.1 & 8.7$\pm$0.6 \\
	P0214003003 & 58527.1 & 1.0 & 15.8$_{-1.2}^{+1.8}$ & 6.2$\pm$0.1 & 0.7$_{-0.2}^{+0.3}$ & 8.3$_{-1.0}^{+1.1}$ \\
	P0214003004 & 58528.4 & 3.2 & 16.3$_{-0.6}^{+0.7}$ & 5.5$\pm$0.1 & 1.2$\pm$0.3 & 8.9$_{-0.7}^{+0.8}$ \\
	P0214003005 & 58529.4 & 3.5 & 17.8$_{-0.7}^{+0.6}$ & 6.1$\pm$0.04 & 0.6$\pm$0.1 & 8.9$\pm$0.5 \\
	P0214003006 & 58530.4 & 3.5 & 16.0$_{-0.6}^{+0.5}$ & 5.3$\pm$0.02 & 0.4$\pm$0.1 & 8.2$\pm$0.5 \\
	P0214003007 & 58533.5 & 5.9 & 16.2$_{-0.5}^{+0.4}$ & 4.9$_{-0.03}^{+0.04}$ & 0.6$\pm$0.1 & 8.2$\pm$0.4 \\
	P0214003008 & 58534.9 & 4.8 & 16.3$\pm$0.6 & 6.4$\pm$0.1 & 1.3$\pm$0.3 & 8.9$_{-0.8}^{+0.3}$ \\
	P0214003009 & 58536.0 & - & - & - & - & - \\
	P0214003010 & 58538.6 & 4.3 & 15.0$_{-0.8}^{+0.9}$ & 5.6$\pm$0.1 & 1.0$_{-0.2}^{+0.3}$ & 8.1$_{-0.7}^{+0.8}$ \\
	P0214003011 & 58541.2 & 4.3 & 17.0$_{-0.5}^{+0.4}$ & 6.4$\pm$0.1 & 1.6$\pm$0.05 & 8.3$_{-0.8}^{+0.9}$ \\
	\enddata
	\tablecomments{Dash lines (-) represent empty GTIs.}
\end{deluxetable*}

\clearpage

\end{document}